\documentclass[preprint,amsmath,amssymb,showpacs,aps,prb]{revtex4}

\usepackage{graphicx}
\usepackage{dcolumn}
\usepackage{bm}

\begin{document}

\title{Nonfrustrated magnetoelectric with incommensurate magnetic order in magnetic field}

\author{A. V. Syromyatnikov}
 \email{syromyat@thd.pnpi.spb.ru}
\affiliation{Petersburg Nuclear Physics Institute, Gatchina, St.\ Petersburg 188300, Russia}

\date{\today}

\begin{abstract}

We discuss a model nonfrustrated magnetoelectric in which strong enough magnetoelectric coupling produces incommensurate magnetic order leading to ferroelectricity. Properties of the magnetoelectric in magnetic field directed perpendicular to wave vector describing the spin helix are considered in detail. Analysis of classical energy shows that in contrast to naive expectation the onset of ferroelectricity takes place at a field $H_{c1}$ that is lower than the saturation field $H_{c2}$. One has $H_{c1}=H_{c2}$ at strong enough magnetoelectric coupling. We show that at $H=0$ the ferroelectricity appears at $T=T_{FE}<T_N$. Qualitative discussion of phase diagram in $H-T$ plane is presented within mean field approach.

\end{abstract}


\maketitle

\section{Introduction}

In the past several years, there has been a revival of interest in magnetic ferroelectrics in which magnetic and ferroelectric orders coexist (magnetoelectrics). \cite{fiebig,most-n} Of particular interest now are systems in
which ferroelectric and spiral magnetic order occur simultaneously due to recognition of the role that such materials might play in fabricating novel magnetoelectric (ME) devices. \cite{har-law} A number of such compounds have been obtained recently: $R$MnO$_3$ with $R=$Gd, Tb, Dy; \cite{tb-field,tb-neutron,kim1} $R$Mn$_2$O$_5$ with $R=$Ho, Y, Tb, Dy;  \cite{25-1,25-2,25-3} Ni$_3$V$_2$O$_8$; \cite{ni-th,ni-exp-f} spinel oxides $R$Cr$_2$O$_4$ with $R=$Co, Fe, Mn; \cite{spinel} MnWO$_4$, \cite{mw} etc. In the majority of these materials paraelectric phase with collinear spin structure (sinusoidal spin density wave) appears below N\'eel temperature $T_N$. Upon further cooling transition to the phase takes place at $T=T_{FE}<T_N$ in which  ferroelectric order coexists with incommensurate elliptical (conical in $R$Cr$_2$O$_4$) magnetic spiral. Such ferroelectric phases are stable down to very small temperature in some of them (e.g.\ $R$MnO$_3$ and $R$Cr$_2$O$_4$) while in others there are transitions to collinear paraelectric phases below $T_{FE}$. All experiments point to key role of the noncollinear spin configurations induced by frustrated exchange interactions in producing the electric polarization. \cite{most-n} Due to frustration $T_N$ and $T_{FE}$ are quite small in all compounds found by now with only one exception, Ba$_{0.5}$Sr$_{1.5}$Zn$_2$Fe$_{12}$O$_{22}$, \cite{kim2} in which $T_{FE}$ is greater than room temperature. A "giant" ME effect is observed in these materials lying in very high sensitivity of the electric polarization to the magnetic field: spin-flop transition in magnetic field is accompanied by rotation of the polarization by $90^\circ$ and by anomaly in dielectric constant. Meantime the value of electric polarization was found to be two-three orders smaller than in the typical ferroelectrics and there is no such great influence of electric field on the magnetic properties indicating smallness of ME coupling in these compounds. Then, many efforts are made now to find materials with stronger ME coupling and with higher transition temperatures which can be used in practice.

Phenomenological treatment of the mechanism of magnetoelectric coupling has been proposed basing on Landau expansion and symmetry consideration. \cite{har,mostovoy,har-law,serg} A microscopic mechanism of the ferroelectricity of magnetic origin has recently been proposed in Ref.~\cite{katsura}, which is based on the idea that spin current ${\bf j}_s\propto [{\bf S}_i\times {\bf S}_j]$ is induced between the noncollinear spins that leads to the electric moment ${\bf P} \propto [{\bf e}_{ij}\times {\bf j}_s]$, where ${\bf e}_{ij}$ is the unit vector connected spins $i$ and $j$. This result can be regarded as an inverse effect of the Dzyaloshinskii-Moriya interaction. As a result one can write the effective ME interaction in the following form: \cite{kat-th}
\begin{equation}
\label{me}
V_{ME} = \beta [{\bf U} \times {\bf e}_{ij}] \cdot \left[ {\bf S}_i \times {\bf S}_j \right],
\end{equation}
where $\bf U$ stands for the corresponding ligand displacement. Taking into account the elastic energy, $\gamma U^2/2$, one finds that the proposed mechanism can lead to ferroelectricity as soon as noncollinear spin structure exists. 

Meantime ME coupling (\ref{me}) can produce a spiral incommensurate magnetic order and electric moment even without frustration if $\beta$ is large enough. Really, let us consider two spins and take into account the direct exchange coupling $J$ between them, ME interaction (\ref{me}) and elastic energy $\gamma U^2/2$. Minimization of the total energy with respect to $U$ and $\phi$, angle between spins, gives $U = (\beta/\gamma) S^2 \sin \phi$ and $\sin \phi \cos \phi = \gamma J/(\beta S)^2 \sin \phi$. Meantime ME coupling constant in TbMnO$_3$, one of the best magnetoelectric of this type by now, is estimated \cite{serg} to be $\beta\sim 1$ meV/\AA. Taking into account that the characteristic value of  $\gamma$ is $10^3$ meV/\AA$^2$ and $J\sim 1$ meV we have from the above estimations $\sin \phi  \cos \phi \approx 10^3 \sin \phi$ that leads to $\phi = 0$ and indicates that in magnetoelectric compounds have been found nowadays frustration is really indispensable for appearance of noncollinear magnetic order leading to ferroelectricity. As we mention above, every efforts are made now to find incommensurate magnetoelectrics with larger ferroelectric moment. But strong ME interaction can produce spiral incommensurate magnetic order and ferroelectric moment without frustration. Thus, it is seen from the above consideration of two spins that if $\beta$ were about 30 times larger than in TbMnO$_3$ a nonzero solution for $\phi$ appears. Moreover, one can expect that among magnetoelectric compounds with strong enough $\beta$ to be obtained (as we hope) nonfrustrated ones would have larger transition temperatures than frustrated ones, other things being equal. 

Thus, it would be instructive to discuss nonfrustrated magnets with strong ME coupling of the form (\ref{me}). Such magnet is considered recently in Ref.~\cite{kat-th}, where collective magnetoelectric modes are discussed. In the present paper we discuss a model similar to that of Ref.~\cite{kat-th} focusing on its properties in magnetic field directed perpendicular to wave vector describing the spin helix. Analysis of classical energy presented in Sec.~\ref{clen} shows that in contrast to naive expectation the onset of ferroelectricity takes place at a field $H_{c1}$ that is lower than the saturation field $H_{c2}$. One has $H_{c1}=H_{c2}$ at strong enough $\beta$. We show that at $H=0$ the ferroelectricity appears at $T=T_{FE}<T_N$. Qualitative discussion of phase diagram in $H-T$ plane is presented within mean field approach in Sec.~\ref{ft}. Sec.~\ref{conc} contains our conclusion.

\section{Classical energy}
\label{clen}

\begin{figure}
\centering
\includegraphics[scale=0.4]{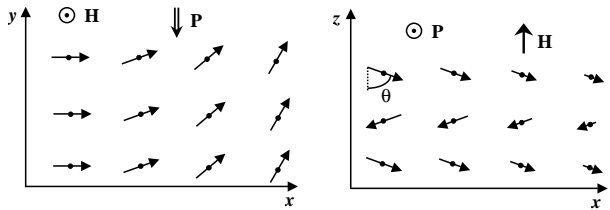}
\caption{Projections of spins on $xy$ and $xz$ planes are shown in the non-collinear phase of magnetoelectric discussed. Magnetic field $\bf H$, electrical polarization $\bf P$ and canting angle $\theta$ of spins in magnetic field are depicted.
\label{schem}} 
\end{figure}

We discuss a magnetoelectric with ferromagnetic interaction $J_{ij}^{xy}$ in $xy$ plane and antiferromagnetic one $J_{ij}^z$ along $z$ axis with ME coupling of the form (\ref{me}) which Hamiltonian has the form
\begin{equation}
\label{ham}
{\cal H} = 
\frac12 \sum_{(i,j)} J_{ij}^{z} {\bf S}_i {\bf S}_j
-
\frac12 \sum_{\langle i,j\rangle} J_{ij}^{xy} {\bf S}_i {\bf S}_j
+
\beta \sum_i [{\bf U}_{{\bf R}_i}\times {\bf e}_x] \cdot \left[ {\bf S}_{{\bf R}_i} \times {\bf S}_{{\bf R}_i+{\bf e}_x}\right]
+
\frac\gamma2 \sum_i U_i^2
+
H\sum_i S_i^z,
\end{equation}
where $(i,j)$ and $\langle i,j\rangle$ denote nearest neighbors along $z$ axis and in $xy$ plane, respectively, ${\bf e}_x$ is the unit vector along $x$ axis, lattice constant is taken to be equal to unity, $\beta$ and $\gamma$ are positive constants and the last term is the Zeeman energy in the field directed along $z$ axis (see Fig.~\ref{schem}). At $H=0$ spins lie in $xy$ plane and their rotation describes by wave vector ${\bf q} = (q,0,0)$. There is a uniform displacement along $y$ axis ${\bf U}_i = {\bf U} = (0,U,0)$. The electric polarization of the sample $\bf P$ is proportional to $N \bf U$, where $N$ is the number of spins in the lattice. When $H\ne0$ the spins cant in opposition to the field direction and makes an angle $\theta<\pi/2$ with $z$ axis (see Fig.~\ref{schem}). Magnetoelectrics with ferromagnetic exchange along $z$ axis or antiferromagnetic exchange in $xy$ plane can be considered on the equal footing. We discuss corresponding results qualitatively in Sec.~\ref{conc}. 

To find $q$, $U$ and $\theta$ one has to minimize the classical energy according to $U$, $q$ and $\theta$ that has the form
\begin{equation}
\label{en}
\frac EN = 2J^zS^2\cos^2\theta - J^{xy}S^2 (\cos^2\theta + \sin^2\theta\cos q) - \beta S^2 U \sin^2\theta\sin q + \frac \gamma2 U^2 - HS\cos\theta,
\end{equation}
where $S$ is the spin value. Eq.~(\ref{en}) has two solutions: (i) that with collinear spin structure, $q=U=0$, and (ii) that with spiral spin structure, $q\ne0$ and $U\ne0$. The last one has the form
\begin{subequations}
\label{res1}
\begin{eqnarray}
U &=& \frac{\beta S^2}{\gamma} \sin^2\theta\sin q,\\
j_{xy} &=& \cos q\sin^2\theta,\\
H &=& \frac{2\beta^2S^3}{\gamma} \cos\theta\left(\sin^2\theta + 2j_z - j_{xy} \right),
\end{eqnarray}
\end{subequations}
where two dimensionless constants are introduced
\begin{equation}
\label{j}
j_{xy} = \frac{\gamma J^{xy}}{(\beta S)^2} \quad \mbox{ and  } \quad j_z = \frac{\gamma J^z}{(\beta S)^2}.
\end{equation}
Stability conditions of the solutions are determined from the demand of positive definiteness of the bilinear form $\partial^2 E/(\partial x \partial y)$, where $x,y=U,q,\theta$. In particular the stability criteria of (\ref{res1}) are given by
\begin{subequations}
\begin{eqnarray}
\label{con30}
q &\ne& 0,\\
\label{con3}
\cos\theta &<& \sqrt{\frac{1 - j_{xy} + 2j_z}{3}}.
\end{eqnarray}
\end{subequations}
It is seen from Eq.~(\ref{res1}b) that this solution exists if
\begin{equation}
\label{con1}
j_{xy} < 1.
\end{equation}
We assume below that condition (\ref{con1}) holds. The second solution of Eq.~(\ref{en}) gives collinear spin structure:
\begin{subequations}
\label{res2}
\begin{eqnarray}
U &=& q = 0 ,\\
\cos\theta &=& 
\left\{
\begin{array}{ll}
H/H_{c2}, & \mbox{ if } H \le H_{c2},\\
1, & \mbox{ if } H > H_{c2},
\end{array}
\right.
\end{eqnarray}
\end{subequations}
where $H_{c2} = 4SJ^z$. Solution (\ref{res2}) is stable at large enough fields so that the following condition satisfies: 
\begin{equation}
\label{con2}
\cos\theta > \sqrt{1-j_{xy}}.
\end{equation}
One concludes from Eqs.~(\ref{con1}) and (\ref{con2}) that if (\ref{con1}) does not hold only collinear spin structure exists. In contrast, when (\ref{con1}) is satisfied, the collinear solution is stable only at $H > H_{c1}^{so}$, where
\begin{equation}
\label{hc1s}
H_{c1}^{so} = 4SJ^z \sqrt{1 - j_{xy}}
\end{equation}
that is found using Eq.~(\ref{res2}b) and assuming the equality in Eq.~(\ref{con2}). The angle $\theta_{c1}^{so}$ corresponding to the field $H_{c1}^{so}$ is given by
\begin{equation}
\label{an1f}
\cos\theta_{c1}^{so} = \sqrt{1-j_{xy}}.
\end{equation}
Notice that at $\theta=\theta_{c1}^{so}$ Eq.~(\ref{res1}b) gives $q=0$.

\begin{figure}
\centering
\includegraphics[scale=0.4]{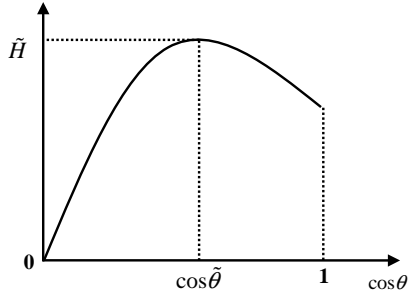}
\caption{Sketch of the right part of Eq.~(\ref{res1}c) as a function of $\cos\theta$. Values of $\tilde H$ and $\tilde\theta$ are given by Eqs.~(\ref{ht}) and (\ref{tt}), respectively.
\label{h}} 
\end{figure}

Let us turn to the transition between spiral and collinear configurations. At $H=0$ and $\theta = \pi/2$ spiral configuration is realized. As it is clear from Eq.~(\ref{res1}c) the angle $\theta$ is an ambiguous function of $H$. A sketch of the right part of Eq.~(\ref{res1}c) is presented in Fig.~\ref{h}. It is seen that there is a maximum at $\theta = \tilde\theta$, where 
\begin{equation}
\label{tt}
\cos\tilde\theta = \sqrt{\frac{1-j_{xy}+2j_z}{3}}. 
\end{equation}
If the right part in Eq.~(\ref{tt}) is larger than unity, it indicates that $\tilde\theta=0$ in consideration presented below and the regime with $\theta > \tilde\theta$ remains only. According to Eq.~(\ref{con3}) one should take those solutions of Eq.~(\ref{res1}c) which has $\theta>\tilde\theta$. The value of magnetic field corresponding to the maximum in Fig.~\ref{h} is given by 
\begin{equation}
\label{ht}
\tilde H = \frac{4\beta^2S^3}{\gamma} \left(\frac{1+2j_z-j_{xy}}{3}\right)^{3/2}. 
\end{equation}
It can be easily shown using Eqs.~(\ref{hc1s}), (\ref{ht}) and Cauchy's inequality that $\tilde H \ge H_{c1}^{so}$. Then, analysis shows that the type of phase transition is determined by the value of the angle $\theta_{c1}^{so}$ at which inequality (\ref{con2}) turns into equality. Two regimes are possible at which we have continuous and discontinuous transitions, respectively: $\theta_{c1}^{so}<\tilde\theta$ and  $\theta_{c1}^{so}>\tilde\theta$. It can be easily shown using Eqs.~(\ref{an1f}) and (\ref{tt}) that $\theta_{c1}^{so}<(>)\tilde\theta$ is equivalent to $j_{xy} + j_z <(>) 1$. Let us discuss these two regimes separately.

\subsection{Continuous transition}

The transition is continuous if $\theta_{c1}^{so}>\tilde\theta$, i.e., if
\begin{equation}
\label{con}
j_{xy} + j_z > 1.
\end{equation}
The angle $\theta$ lowers as the field rises and the spiral solution turns into collinear one at $H=H_{c1}^{so}$: $q$ and $U$ reduce gradually to zero as $H$ approaches $H_{c1}^{so}$ and $\theta$ approaches $\theta_{c1}^{so}$; at $H=H_{c1}^{so}$ we have $\theta = \theta_{c1}^{so}$, $q=U=0$, the stability criterion of the spiral solution (\ref{con30}) ceases to hold and the collinear solution (\ref{res2}) becomes stable (criterion (\ref{con2}) begin to hold). All spins become parallel the field direction at $H=H_{c2}$. As a result we obtain phase diagram shown in Fig.~\ref{phase}(a) corresponding to the line $T=0$.

\subsection{Discontinuous transition}

\begin{figure}
\centering
\includegraphics[scale=1.0]{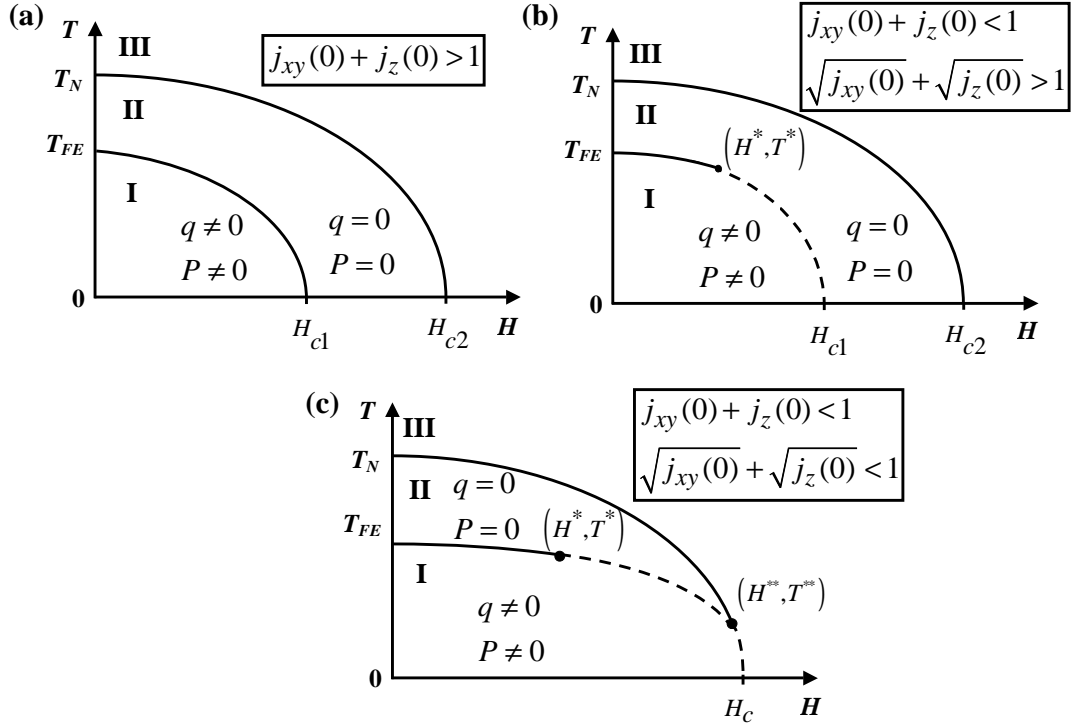}
\caption{Phase diagram in the $H-T$ plane of the magnetoelectric discussed in some limiting cases indicated within frame in each plane (constants $j$ are given by Eqs.~(\ref{j}) and $j(0) = j(T=0)$). There is an incommensurate magnetic order with wave vector $\bf q$ and electric moment $\bf P$ in phase I. Spins are parallel each other and canted by a finite angle to the magnetic field in paraelectric phase II. All spins are parallel to the magnetic field and $P=0$ in phase III. Solid and dashed lines denote lines of second and first order phase transitions, respectively. One goes on successively from plane (a) to (b) and (c) increasing $\beta$. (a) The line of phase transitions between I and II is given by Eq.~(\ref{hc1t}) within MFA. (b) Transitions between I and II is of the second and of the first order above and below $T^*$, respectively, where $T^*$ is given by Eq.~(\ref{t*}). The line of phase transitions in this case is given within MFA by Eqs.~(\ref{hc1t}) and (\ref{hc1tf}) above and below $T^*$, respectively. (c) There is only one critical field $H_c$ at $T=0$ in this regime. Temperatures $T^*$ and $T^{**}$ are given by Eqs.~(\ref{t*}) and (\ref{t**}), respectively.
\label{phase}} 
\end{figure}

The transition is discontinuous if $\theta_{c1}^{so}<\tilde\theta$, i.e., if
\begin{equation}
\label{dis}
j_{xy} + j_z < 1.
\end{equation}
This regime corresponds to larger spin-lattice coupling (larger $\beta$) than that discussed above. In this case the angle $\theta_{c1}$ can not be reached gradually because, in particular, the spiral solution is unstable at $\theta<\tilde\theta>\theta_{c1}^{so}$. Thus the transition is of the first order in this case. When $H$ reaches $H_{c1}^{so}$, $q$ does not turn into zero and the spiral solution remains stable. At the same time the collinear solution is also stable at $H>H_{c1}^{so}$ but the energy of spiral solution is lower than that of collinear one at $H=H_{c1}^{so}$. As one increases the magnetic field further, the ground state energies of these two solutions comes together and the transition takes place when they become equal. Corresponding field can be greater or lower than $H_{c2}$. In the first case all spins in the collinear phase are parallel to the field ($\theta_{c1}^{fo}=0$) whereas in the second case $\theta_{c1}^{fo}\ne0$ and all spins becomes parallel each other at $H>H_{c2}$ only. It is easy to find in the second scenario using Eqs.~(\ref{en}), (\ref{res1}) and (\ref{res2}) for the critical field 
\begin{equation}
\label{hc1f}
H_{c1}^{fo} = 2SJ^z \frac{1-j_{xy}+j_z}{\sqrt{j_z}}.
\end{equation}
The transition is accompanied by the drop of the angle $\theta$ from $\theta_{c1}^{fo}$ given by
\begin{equation}
\label{thf}
\cos\theta_{c1}^{fo} = \sqrt{j_z}
\end{equation}
to $\theta_{c1}^{so}$ given by Eq.~(\ref{an1f}) and the wave vector $q$ of the spiral switches from
\begin{equation}
\cos q_{c1}^{fo} = \frac{j_{xy}}{1-j_z}
\end{equation}
to $q=0$. Then we lead to the part of phase diagram in $H-T$ plane shown in Fig.~\ref{phase}(b) corresponding to the line $T=0$. 

One obtain that $H_{c1}^{fo}<H_{c2}$ if 
\begin{equation}
\label{con4}
\sqrt{j_{xy}} + \sqrt{j_z} > 1.
\end{equation}
If $\beta$ is as large as this criterion does not hold, there is only one critical field $H_c$ and we lead to the line $T=0$ on the phase diagram shown in Fig.~\ref{phase}(c). Expression for $H_c$ is quite complex and we do not present it here.

Let us discuss now the phase diagrams at $T>0$. 

\section{Finite temperatures}
\label{ft}

Let us find the equation for the line of phase transitions between collinear and spiral phases within the mean-field approximation (MFA). The energy (\ref{en}) is a function of $T$. Working in MFA we imply that the spin value is reduced by thermal fluctuations:
\begin{equation}
\label{s}
S(T) = S\left( 1 - \frac{T}{T_N} \right),
\end{equation}
where $T_N = 2S^2(J^z + 2J^{xy})$ within MFA, $S\gg1$ and $T\ll T_N$. For simplicity we neglect dependence of constants $\beta$, $\gamma$ and $J$ on $T$ in Eq.~(\ref{en}). As soon as the spin value depends on $T$, all quantities depending on $S$ arising in the above discussion are also functions of $T$. In particular, using Eqs.~(\ref{hc1s}) and (\ref{hc1f}) one can find the line of phase transitions in $H-T$ plane between spiral and collinear phases. Meantime this line is different in the case of second and first order phase transitions at $T=0$, i.e., at $j_{xy}(T=0) + j_z(T=0) > 1$ and $j_{xy}(T=0) + j_z(T=0) < 1$, that should be discussed separately.

\subsection{Second order phase transition at $T=0$ (i.e., $j_{xy}(T=0) + j_z(T=0) > 1$)}

Evidently, the line of phase transition started at $(H=0,T_N)$ should lead to the point $(H_{c2},T=0)$. It is also clear that the line of phase transition between non-collinear and collinear phases should start at $(H_{c1},0)$ and end at $(0,T_{FE})$, where $T_{FE}<T_N$. Using Eqs.~(\ref{hc1s}) and (\ref{s}) one can easily find the equation on $H_{c1}^{so}(T)$:
\begin{equation}
\label{hc1t}
\frac{T}{T_N} = \frac{ [H_{c1}^{so}(0) ]^2 - [H_{c1}^{so}(T)]^2}{2(4SJ^z)^2}
\end{equation}
that gives a parabola in $H-T$ plane (see Fig.~\ref{phase}(a)). In particular, we have from Eq.~(\ref{hc1t}) $T_{FE} = T_N[1-j_{xy}(0)]/2$. Then the demand $T\ll T_N$ implies that $j_{xy}(0)\sim1$.

\subsection{First order phase transition at $T=0$ (i.e., $j_{xy}(T=0) + j_z(T=0) < 1$)}

The phase diagram is different depending on whether condition (\ref{con4}) holds or does not hold at $T=0$. Let us start with the first case.

\subsubsection{$\sqrt{j_{xy}(0)} + \sqrt{j_z(0)} > 1$}

It is seen from Eqs.~(\ref{j}) that thermal fluctuations increase $j_{xy}$ and $j_z$. Then, above a certain temperature $T^*$ the sum $j_{xy}(T) + j_z(T)$ becomes larger than unity. Hence, at $T>T^*$ and $T<T^*$, where
\begin{equation}
\label{t*}
T^* = T_N \frac{1 - j_{xy}(0) - j_z(0)}{2}
\end{equation}
that is found from the condition $j_{xy}(T^*) + j_z(T^*) = 1$, the transition is of the second and first orders, respectively. The line of the second order phase transitions at $T>T^*$ is given by Eq.~(\ref{hc1t}). The line of the first order phase transitions at $T<T^*$ can be found using Eq.~(\ref{hc1f}) with the result
\begin{equation}
\label{hc1tf}
\frac{T}{T_N} = \frac{H_{c1}^{fo}(0) - H_{c1}^{fo}(T)}{4SJ^z}\sqrt{j_z(0)}.
\end{equation}
Then, we lead to plane (b) in Fig.~\ref{phase}.

\subsubsection{$\sqrt{j_{xy}(0)} + \sqrt{j_z(0)} < 1$}

Phase diagram in this case is presented in Fig.~\ref{phase}(c). The temperature $T^{**}$ is found from the condition $\sqrt{j_{xy}(T^{**})} + \sqrt{j_z(T^{**})} = 1$ with the result 
\begin{equation}
\label{t**}
T^{**} = T_N \left(1 - \sqrt{ j_{xy}(0) } - \sqrt{ j_{z}(0) }\right).
\end{equation}
At $T<T^{**}$ there are transitions from spiral phase to collinear one with all spins are along the field direction (phase III). In contrast, at $T>T^{**}$ there is first a transition to collinear phase with $\theta\ne0$ and then second order phase transition to the phase III. The phase transition between spiral and collinear phases is of the first and second orders at $T<T^*$ and $T>T^*$, respectively, where $T^*$ is given by Eq.~(\ref{t*}). 

\section{Conclusion}
\label{conc}

We discuss in the present paper a nonfrustrated magnetoelectric (\ref{ham}) in magnetic field with spin-lattice coupling of the form (\ref{me}) that is strong enough to produce spiral spin structure inducing ferroelectricity. Ground state energy is analyzed. We show that, in contrast to naive expectation, the onset of ferroelectricity takes place at $H<H_{c1}$ and $H_{c1}$ is lower than the saturation field $H_{c2}$ if constant $\beta$ in Eq.~(\ref{me}) is not too large. The type of the phase transition between collinear paraelectric phase and spiral ferroelectric one depends on values of constants $j$ given by Eqs.~(\ref{j}): the transition is of the second and first order if condition (\ref{con}) holds and does not hold, respectively. Moreover, if inequality (\ref{con4}) does not satisfy, there is only one critical field at which transition of the first order takes place from spiral phase to that in which all spins are parallel to the field.

As a result of qualitative consideration using mean field approach we obtain phase diagram in $H-T$ plane shown in Fig.~\ref{phase}. One goes on successively from plane (a) to (b) and (c) increasing $\beta$. It should be noted also that the phase diagram remains qualitatively the same for a magnetoelectric with antiferromagnetic coupling among spins in $xy$-plane (cf.\ Eq.~(\ref{ham})). In contrast, if the exchange coupling along $z$ axis is ferromagnetic, the constant $j_z$ should be put equal to zero in the above consideration. As a consequence one leads to phase diagram shown in Fig.~\ref{phase}(c).

\begin{acknowledgments}
This work was supported by Russian Science Support Foundation, President of Russian Federation (grant MK-4160.2006.2), RFBR (grants 06-02-16702 and 06-02-81029) and Russian Programs "Quantum Macrophysics", "Strongly correlated electrons in semiconductors, metals, superconductors and magnetic materials" and "Neutron Research of Solids".
\end{acknowledgments}

\newpage


\end{document}